%% file: main.tex
\documentclass[conference]{IEEEtran}
\IEEEoverridecommandlockouts

\usepackage{epsfig,xspace,url}
\usepackage{amssymb,amsmath,latexsym,amsfonts,amsthm}
\usepackage{graphicx}
\usepackage{bm}
\usepackage{mathtools}
\usepackage{enumitem}
\usepackage{multicol}
\usepackage{booktabs} 
\usepackage{algorithm}
\usepackage{subcaption}
\usepackage[noend]{algpseudocode}
\usepackage{balance}
\usepackage{multirow}
\usepackage{colortbl}

\usepackage{cite}
\usepackage{textcomp}
\usepackage{xcolor}
\def\BibTeX{{\rm B\kern-.05em{\sc i\kern-.025em b}\kern-.08em
    T\kern-.1667em\lower.7ex\hbox{E}\kern-.125emX}}

\makeatletter
\DeclareRobustCommand*\cal{\@fontswitch\relax\mathcal}
\makeatother

\PassOptionsToPackage{hyphens}{url}\usepackage{hyperref}

\usepackage{color}

\usepackage{nomencl}
\makenomenclature

\begin{document}

\title{Quantifying an Interference-Assisted Signal Strength Breathing Surveillance Attack}

\input{authors}
\maketitle

\input{abstract}

\input{keywords}

\input{intro}
\input{threat_model}
\input{background}
\input{experiment}

\input{crb}

\input{mitigation}

\input{related}

\input{discussion}

\input{conclusion}
\input{ack}
\balance

\bibliographystyle{IEEEtran}
\input{ref.bbl}

\end{document}

%% file: authors.tex
\author{\IEEEauthorblockN{Alemayehu Solomon Abrar\IEEEauthorrefmark{1},
Neal Patwari\IEEEauthorrefmark{1}\IEEEauthorrefmark{2}\IEEEauthorrefmark{4}, Aniqua Baset\IEEEauthorrefmark{3} and
Sneha Kumar Kasera\IEEEauthorrefmark{3}}
\IEEEauthorblockA{
\IEEEauthorrefmark{1} \textit{Dept. of Electrical \& Systems Engineering}, \textit{Washington University in St. Louis}, St. Louis, USA \\
\IEEEauthorrefmark{2} \textit{Dept. of Computer Science \& Engineering}, \textit{Washington University in St. Louis}, St. Louis, USA \\
\IEEEauthorrefmark{3} \textit{School of Computing}, \textit{University of Utah}, Salt Lake City, USA \\
\IEEEauthorrefmark{4} \textit{Xandem}, Salt Lake City, USA \\
Email: alemayehusolomon@wustl.edu, npatwari@wustl.edu, aniqua@cs.utah.edu, kasera@cs.utah.edu }}

%% file: abstract.tex
\begin{abstract} 
A malicious attacker could, by taking control of internet-of-things devices, use them to capture received signal strength (RSS) measurements and perform surveillance on a person's vital signs, activities, audio in their environment, and other RF sensing capabilities.  This paper considers an attacker who looks for periodic changes in the RSS in order to surveil a person's breathing rate.  The challenge to the attacker is that a person's breathing causes a low amplitude change in RSS, and RSS is typically quantized with a significantly larger step size.  This paper contributes a lower bound on an attacker's breathing monitoring performance as a function of the RSS step size and sampling frequency so that a designer can understand their relationship.  Our bound considers the little-known and counter-intuitive fact that an adversary can improve their sinusoidal parameter estimates by making some devices transmit to add interference power into the RSS measurements.  We demonstrate this capability experimentally.  As we show, for typical transceivers, the RSS surveillance attack can monitor RSS with remarkable accuracy. New mitigation strategies will be required to prevent RSS surveillance attacks.

\end{abstract}

%% file: keywords.tex
 \begin{IEEEkeywords}
 Received signal strength, respiratory rate monitoring, privacy of personal health information
  \end{IEEEkeywords}

%% file: intro.tex
\section{Introduction}
\label{sec:intro}

Internet-of-things (IoT) devices are notoriously easy to compromise \cite{ifsec, Antonakakis2017understanding, reaper}. Given that devices bring sensors like microphones and cameras into our private spaces \cite{warren2017amazon}, people are rightfully concerned for their privacy. People know what kind of information an attacker could obtain from compromising a video camera in their private spaces, and may not deploy them \cite{townsend2011privacy} or may physically disable them \cite{warren2017amazon}. In contrast, there is little awareness of what an attacker could obtain from a device which can measure received signal strength (RSS).  Yet, \emph{every} wireless device could be an RF sensor. 

RF sensors have been shown to be capable of monitoring breathing and heart rate \cite{liu2015tracking}, location \cite{wang2017csi,patwari2010rf}, activity \cite{wang2015understanding}, gesture \cite{pu2013whole}, audio \cite{wei2015acoustic}, and keystrokes \cite{ali2015keystroke}. 
Many of these demonstrations have used channel state information (CSI) which can only be obtained from a select group of WiFi network interface cards (NICs).  From the designer's perspective, an attack using CSI can be negated by avoiding these NICs.  However, wireless devices universally allow access to RSS information since network functions such as multiple access control and power control \cite{baccour2012radio,kim2006cara,lin2006atpc} require it.  Thus, in this paper, we address the most universal problem: the capabilities of an attacker who compromises a wireless device to access RSS measurements and surveil a person near that device.  The use of RSS measurements for sensing is also not new, as they have been used to perform contact-free breathing monitoring \cite{kaltiokallio2014non, ravichandran2015wibreathe}, device-free localization \cite{youssef2007challenges}, and gesture and activity recognition \cite{sigg2014telepathic}.  However, the extent to which an attacker could perform any RF sensing task using RSS is still an open question, one which motivates this paper.

Our purpose is to address the question, what is the best that an attacker could possibly do?  Knowing an attacker's limits can provide a guarantee to a user, that even if the device is completely compromised by an attacker, its ability has a particular quantitative limit.  Furthermore, if these limits are known as a function of transceiver parameters, then a transceiver designer can adapt the design to reduce an attacker's capabilities.  

\begin{figure*}[tbhp]
\centering
   \begin{subfigure}[t]{0.9\textwidth}
     \includegraphics[width=0.45\textwidth]{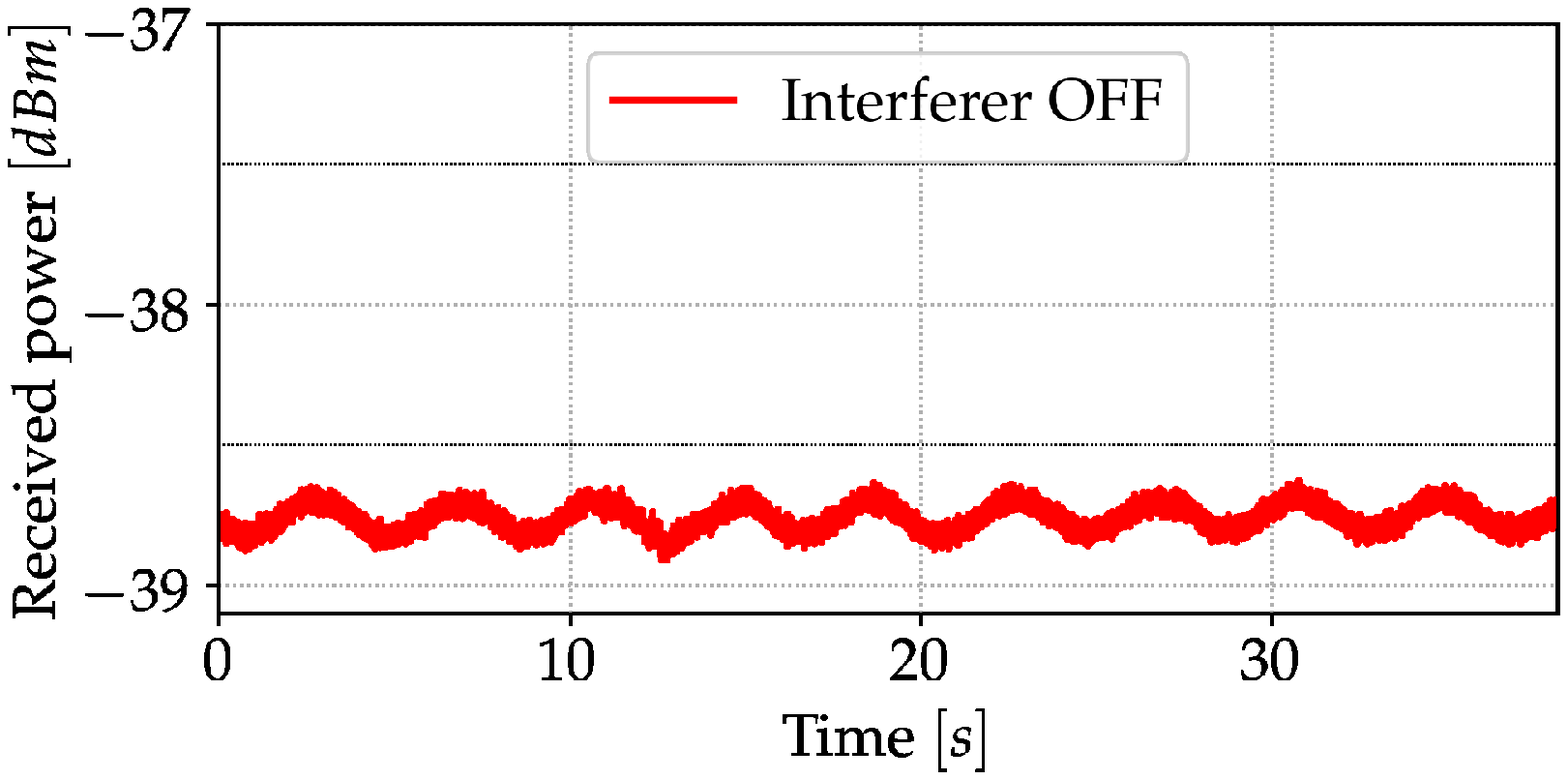}
     \qquad
     \includegraphics[width=0.45\textwidth]{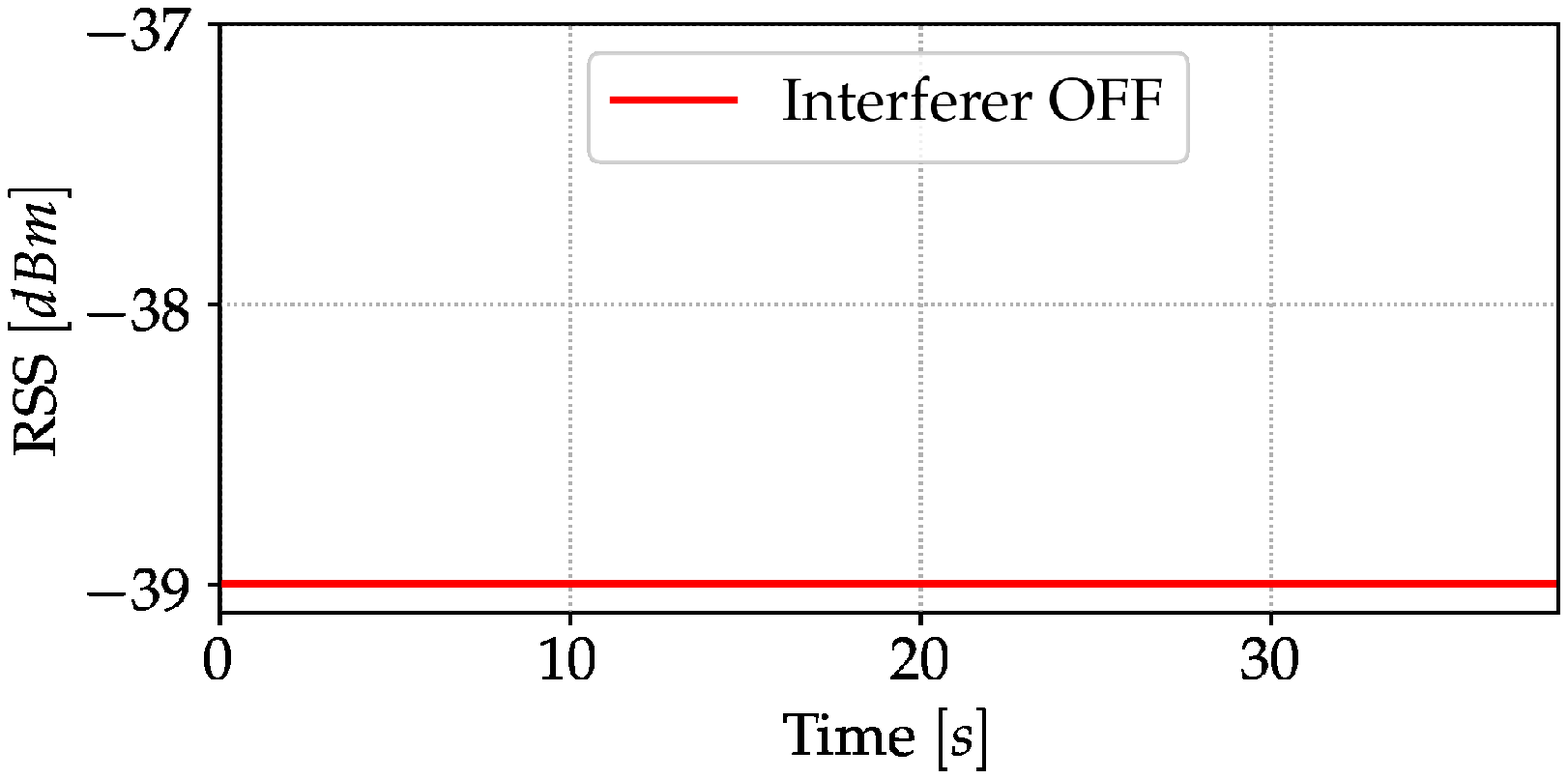}
     \label{fig:intoff}
  \end{subfigure} %
  \begin{subfigure}[t]{0.9\textwidth}
     \includegraphics[width=0.45\textwidth]{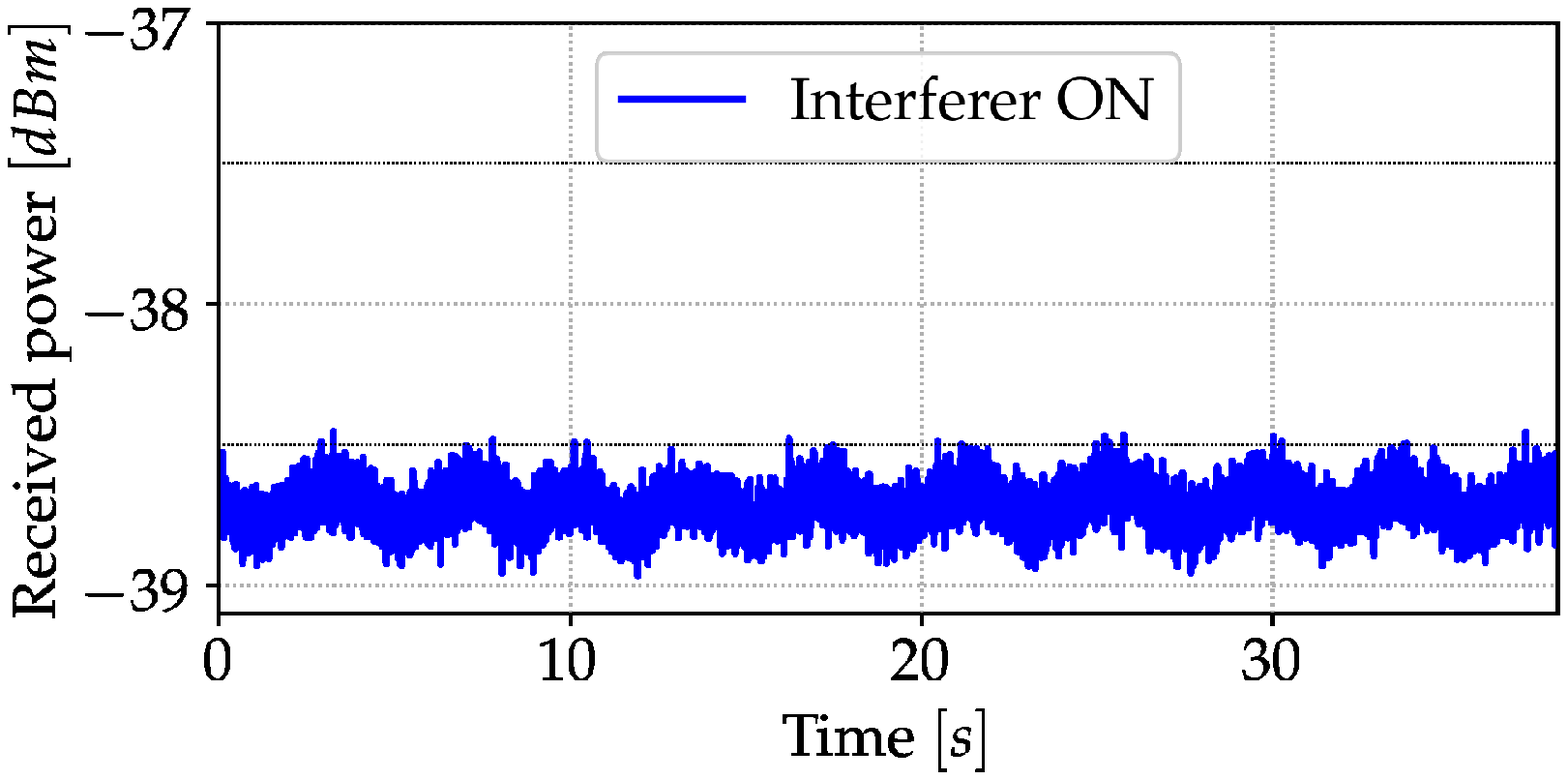}
     \qquad
     \includegraphics[width=0.45\textwidth]{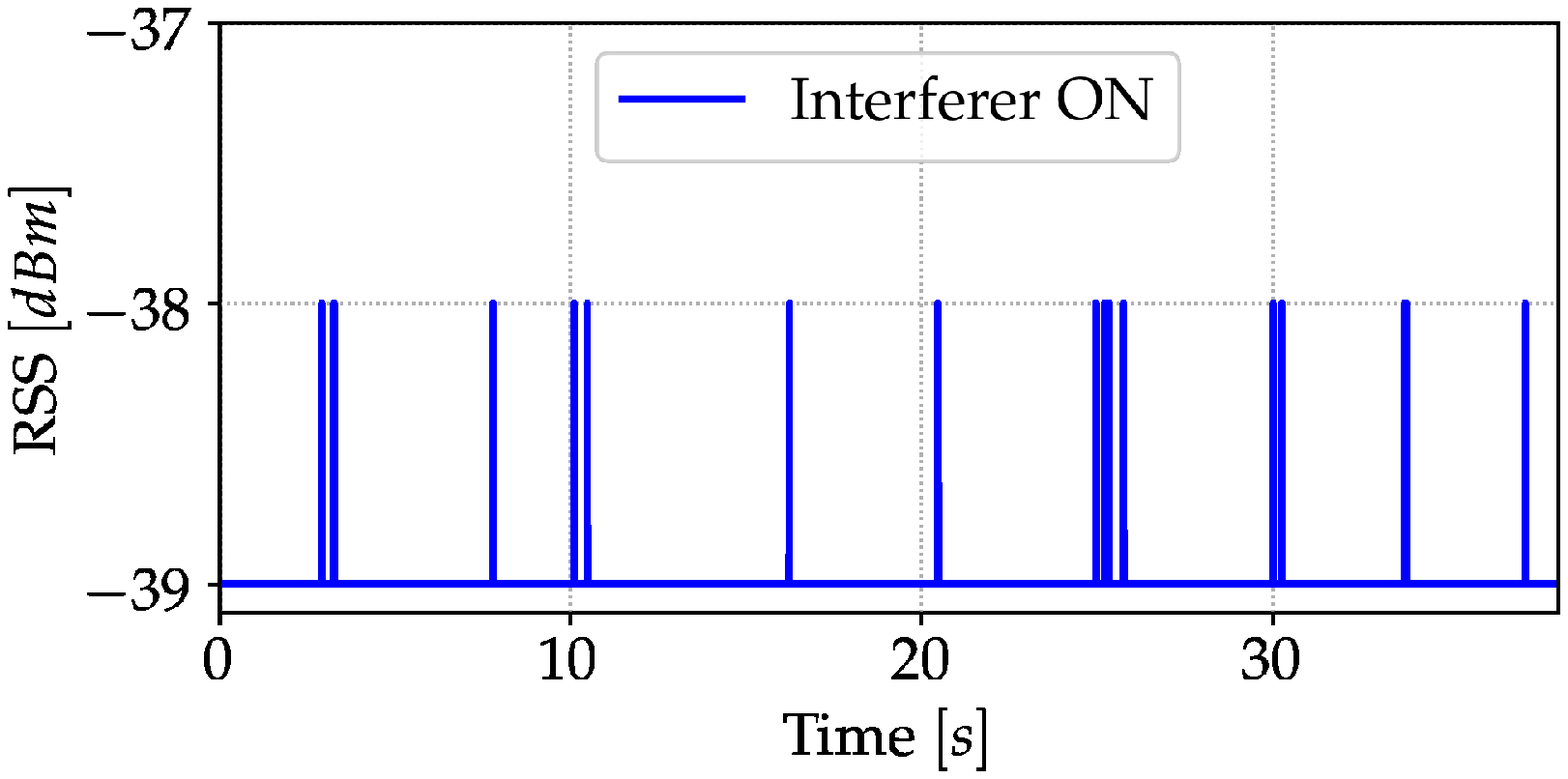}
     \label{fig:inton}
  \end{subfigure}

  \caption{Received power (Left column)  and RSS quantized to the nearest integer (Right column), while a person breathes at 15 bpm.  (Top row) Without interference, RSS is constant, but (Bottom row) with interference the RSS crosses to the higher value at least once per period, enabling estimation of breathing rate. }
 \label{fig:noise_rss}
 \end{figure*}

\vspace{0.1in} \noindent {\bf Quantization and Interference}: Quantization is good news for security against an attacker with access to RSS. A well-known limitation of RSS is that it is quantized, typically with 1 dB step size (although sometimes 0.5 dB or 4 dB). For the above sensing applications, typical changes to received power are much less than 1 dB (e.g., 0.1 dB in \cite{luong2016rss}) unless the user is very close to the transceivers. Fig. \ref{fig:noise_rss}(Top row) shows what often happens---the received power is affected by a person breathing at 15 breaths per minute (bpm), but the quantized RSS is constant.  

However, the bad news we discover in this investigation is that an attacker's capabilities are greater than previously thought because the attacker can exploit what we call \emph{helpful interference} (HI).  HI is the purposeful transmission of interference from other devices to increase the variance of the RSS measurement at a receiver.  An attacker could use other compromised devices to transmit, perhaps with carrier sense disabled, to generate HI.  Counterintuitively, this increase in measurement variance prior to quantization can actually improve the attacker's estimates of frequency and amplitude, partially negating the effects of quantization.  Fig. \ref{fig:noise_rss}(Bottom row) shows an example of how noise from an interferer helps to sometimes ``push'' the quantized RSS over the boundary to the next RSS value, thus making the breathing rate observable.    We provide the first experimental demonstration, to our knowledge, of the ability to use interference to improve the performance of an RSS-based breathing rate estimator. Our experimental observations with varying levels of helpful interference also exhibit an optimal level of HI.  The estimation bounds presented in this paper take into account an attacker's ability to use HI, and also show that there is an optimal level of HI beyond which performance degrades.  The resulting variance bounds are a function of the transceiver's RSS quantization step size and sampling rate.  Device makers can use this bound to limit the inadvertent measurement capabilities of attackers by the design of their device, and we provide strategies for this purpose in this paper.

\vspace{0.1in} \noindent {\bf Breathing Monitoring}:  We pay particular attention to RSS-based breathing monitoring in this paper, as it has already been demonstrated \cite{patwari2014breathfinding}.  We believe it is important to know the relationship between breathing rate estimation performance and quantization step size and RSS sampling rate, which can be computed from the bound in this paper.  Further, RSS breathing rate is private health information.  Resting heart and breathing rates are measures of cardiovascular health.  During sleep, such data is used to classify sleep stage and thus sleep quality.  Vital sign data is useful for knowing emotional state.  Typically a person would not want an employer or an insurance company to have such data.  Together with other data, breathing rate can help to classify a person's activities, including in private spaces like bedrooms.

\vspace{0.1in} \noindent {\bf Contribution Summary}: We summarize the contributions of this paper as follows:
\begin{enumerate}
    \item This paper is the first to quantify and experimentally validate the use of helpful interference, in this case, for breathing monitoring.
    \item This paper provides a lower bound on frequency rate and amplitude estimation, and applies it to provide quantitative results, for RSS-based breathing monitoring.
    In particular, we find the lower bound on the variance of any unbiased estimator of frequency or amplitude of a sinusoid in the RSS signal, using Cram\'er-Rao  bound (CRB) analysis.
    \item Based on the CRB results, we also provide a set of strategies that a transceiver designer can use for providing RSS in a way that prevents accurate breathing rate eavesdropping.
\end{enumerate}
In combination, these contributions provide quantitative limits on the effectiveness of an attacker who compromises wireless devices and uses them as RF sensors.  

%% file: threat_model.tex
\section{Threat Model} 
\label{sec:threat} 
We assume an area in which a person is located in a home with two or more transmitters and receivers.
When the environment is otherwise static, periodic motion caused by the person, for example, due to breathing and pulse, cause changes in the radio channel that can be observed at the receivers as variations in the received power, and indirectly in RSS, which is the quantized received power.

We assume that an attacker can access the transmitters and receivers in the home, and that they can alter device firmware or software to force transmitters to transmit more often and  receivers to receive (and collect RSS measurements) more often, up to the maximum rate and maximum RSS precision as possible with the receiver hardware. 
Since wireless standards (e.g, 802.11) require higher layer access to signal strength \cite{sjoberg2010measuring}, an attacker can use this information maliciously for RSS surveillance.
This attack model is practical as it has been shown that there are millions of vulnerable and unprotected Internet connected devices deployed today, and attackers have repeatedly managed to remotely take over such devices and install botnets on them \cite{Antonakakis2017understanding, reaper} or make modifications to the software/firmware \cite{bazhaniuk2018remotely}. 
Furthermore, we assume that an attacker can force a transmitter to transmit in the same frequency band at the same time as the other transmitter (e.g., by disabling carrier sensing \cite{xu2005feasibility, vanhoef2014advanced}, using a hidden terminal) 
in order to contribute noise to the receiver, as described in \S \ref{sec:eavesdrop}.

The attacker can either transfer the measurements to another processor or process the measurements locally on the same device.  We do not assume any computational or communication constraints for the attacker. We make no assumption about the algorithm used except that it is unbiased, e.g., the attacker does not always guess the same breathing rate regardless of the data.

We do not consider an adversary that brings their own wireless devices to the home. While an attacker who brings a software-defined radio (SDR) to a home might be able to monitor a resident with greater accuracy, this would require physical proximity to each home to be attacked and considerable cost for each SDR.  In contrast, the attack we study requires only remote access to the already installed but compromised commercial wireless devices, and thus could be launched without new hardware and on a very large scale.

%% file: background.tex
\section{Background}
\label{sec:eavesdrop}

With the increasing density of wireless interfaces in our private spaces which grant users and applications direct access to RSS measurements, and the variety of abilities of RF sensing, we anticipate RSS-based ``surveillance'' as a potential threat to privacy.  
In the following subsections, we discuss existing methods for RSS-based respiration monitoring, briefly describing reported algorithms for breathing rate estimation. %

\subsection{RSS Extraction and Pre-processing}

Most existing wireless technologies, such as WiFi and Zigbee, provide direct access to RSS information. In IEEE 802.11 implementations, one may use RSS readings to measure the relative signal power level of the preamble of each received 802.11 frame. To measure RSS, an adversarial user could passively listen for transmitted frames or initiate packet reception by continuously pinging a nearby device. In most existing operating systems, 
RSS is read with a single command.
Many smart phone applications access WiFi signal strength, and there are few limits on software access to RSS measurements.

After collection, further processing is typically performed. 
Filtering is used to remove the DC component and any high-frequency noise, neither of which is informative for breathing rate estimation.
Specifically, for breathing monitoring results in this paper, we use: 1) a DC removal filter applied to each 30-second window RSS measurements, and 2) a $4^{th}$ order Butterworth lowpass filter with a cutoff frequency of 0.5 Hz. 

\begin{figure}[tbhp]
\begin{minipage}[b]{1.0\linewidth}
\centerline{\includegraphics[width=0.9\columnwidth]{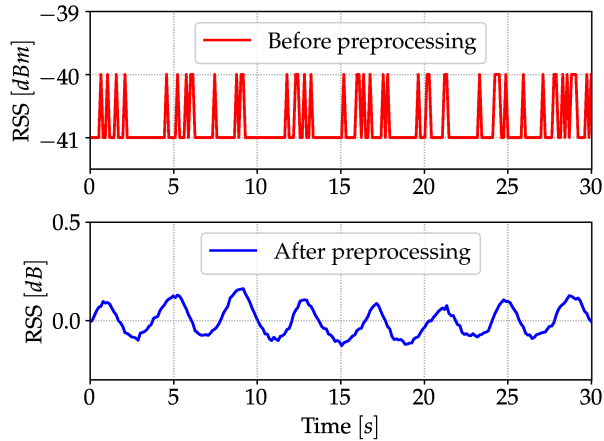}}
\end{minipage}
\caption{RSS pre-processing by DC removal \& filtering}
\label{fig:preprocess}
\end{figure}

We show in Fig. \ref{fig:preprocess} RSS data  recorded from a subject lying on a cot and breathing at 15 bpm. After the DC removal filtering is applied to the RSS measurement, one can see about seven and half cycles over 30 seconds, which corresponds to a breathing rate of 15 bpm.

\subsection{Respiratory Rate Estimation}

\begin{figure}[tbph]
\begin{minipage}[b]{1.0\linewidth}
\centerline{\includegraphics[width=0.8\columnwidth]{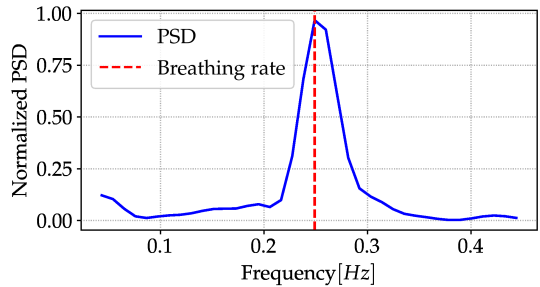}}
\end{minipage}
\caption{MLE of breathing rate for data in Fig. \ref{fig:preprocess}}
\label{fig:psd}
\end{figure}

Next, the pre-processed RSS data is used to estimate breathing rate. There are various methods proposed for frequency estimation, such as zero-crossing detection, Fourier transform maximum selection, linear predictive coding, and least-squares harmonic analysis \cite{kootsookos1993review, ravichandran2015wibreathe}. We apply a maximum likelihood estimator (MLE) as it provides generally unbiased and efficient estimation.
The MLE algorithm finds the peak frequency $\hat{f}$ in the power spectral density (PSD) within the respiratory rate range $[f_{min}, f_{max}]$. As shown in Fig. \ref{fig:psd}, the peak frequency $\hat{f}$ of the PSD matches the actual breathing rate.

The amplitude of the breathing-induced signal is also useful.  It can be applied in breathing detection or to find a breathing person, e.g., in a collapsed building.  One algorithm is to use the amplitude of the PSD at the breathing frequency $\hat{f}$.

%% file: experiment.tex
\section{Breathing Surveillance with Helpful Interference}
\label{sec:eval}

In this section, we describe and demonstrate RSS-based breathing surveillance with helpful interference. To the best of our knowledge, we are the first to introduce the use of helpful interference, that is, transmitting interference to overcome the limitation of RSS quantization in estimating breathing rate.

\subsection{Devices and Setup}

For our evaluation, we desire a commercial wireless transceiver, but we need control over the quantization step size and sampling rate. We achieve this goal by using a commercial wireless transceiver, the TI CC1200. The CC1200 radio is used as integral part of some commercial internet-of-things (IOT) products like the zoul module \cite{zoul} and it has been shown to measure received power without quantization \cite{luong2016rss}.
We then can apply any quantizer to the received power in post-processing to emulate the RSS that would have been reported using an arbitrary transceiver RSS quantization step size.

We use CC1200 transceivers configured as transmitter, receiver and interferer nodes respectively.  The transceivers are configured to operate over a less congested ISM band at 868/915 MHz to have a better control on the level of interference in the channel.  While we believe that uncontrolled interference could also benefit breathing monitoring, we control our interference source for purposes of understanding the relationship between interference power and monitoring performance.  For simplicity, the transmitter sends a continuous wave signal at a transmit power of 12 dBm. The receiver node uses the average of the squared magnitude of the complex baseband (IQ) samples to compute the received power, and outputs a received power measurement at a rate of 615 Hz. The third transceiver is programmed to generate HI. We implement a 2-FSK transmitter with a symbol rate of 256 Kbps in the same band, and transmit random symbols. To study the effect of the magnitude of interference, we also control the output power of the interferer by modifying the value of the \texttt{PA\_CFG1} register on the CC1200 transceiver.

\begin{figure}[t]
\begin{minipage}[b]{1.0\linewidth}
\centerline{\includegraphics[width=0.95\columnwidth]{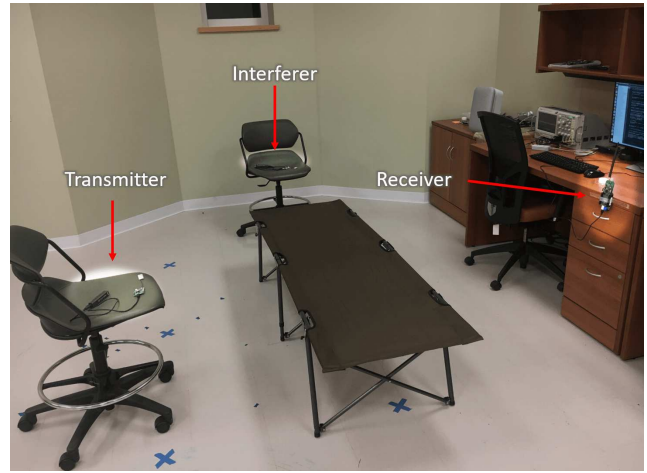}}
\end{minipage}
\caption{Experiment Setup}
\label{fig:setup}
\end{figure}

The experiments are conducted in a laboratory setting in a university building. We run the experiments with three CC1200 transceivers operating as a transmitter and a receiver, and a helpful interference transmitter (HI TX). 
Fig. \ref{fig:setup} shows one setup used in this experiment.  All experiments are performed with a single user lying on a cot. The subject inhales and exhales at a fixed rate given by a metronome. 

We use root mean squared error (RMSE) as error metric to evaluate the breathing monitoring performance.

\subsection{Helpful Interference}
\begin{figure*}[tbhp]
  \begin{subfigure}[t]{0.48\textwidth}
     \includegraphics[width=\textwidth]{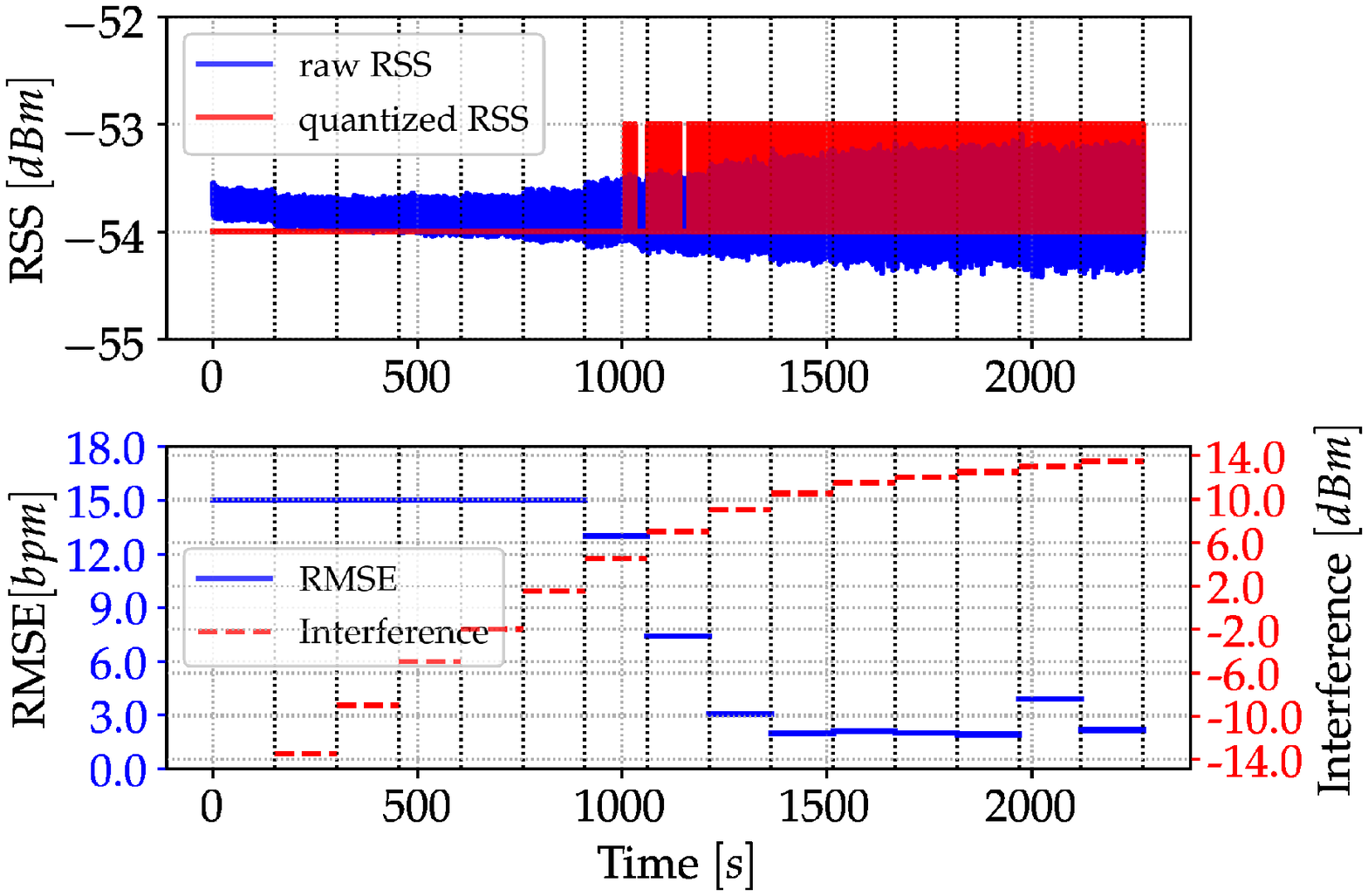}
    \end{subfigure}
  \hfill
  \begin{subfigure}[t]{0.45\textwidth}
     \includegraphics[width=\textwidth]{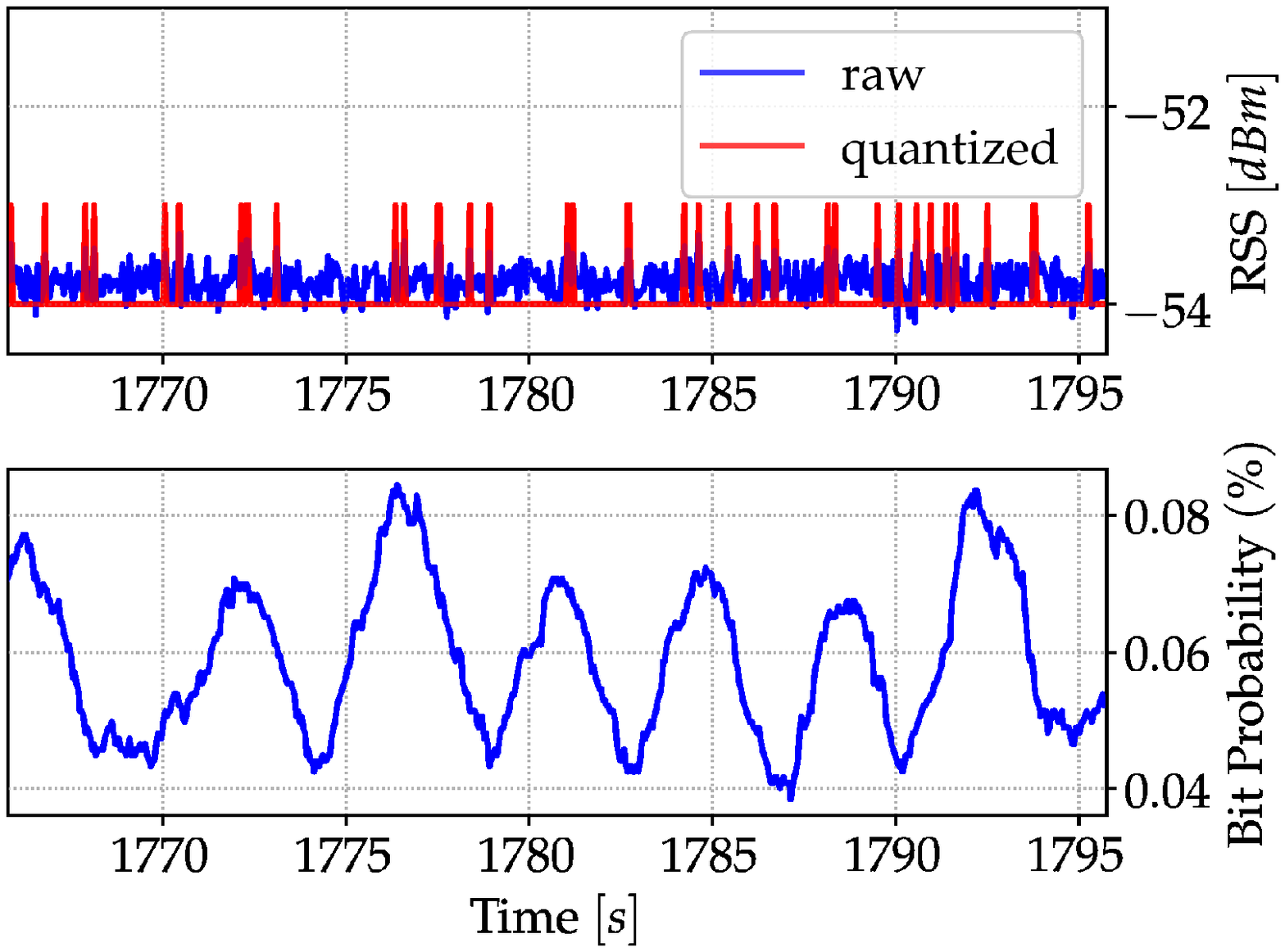}
     \label{fig:raw_vs_zoom}
  \end{subfigure} %
\caption{As HI power gain increases each 151s, the breathing rate error decreases to a minimum of 2 bpm.}
\label{fig:interference}
\end{figure*}

First, we evaluate the effect of increasing interference power on breathing rate estimation using quantized RSS measurements. In Fig. \ref{fig:interference}, we show how RSS changes in the presence of increasing interference.  With no interference at the start of the experiment, the measured RSS almost always takes the value of -54 dBm.  As a result, it is largely unable to estimate breathing rate, and the breathing rate RMSE is about 15 bpm.  Each 151 seconds, the HI power is increased by by changing the value of the \texttt{PA\_CFG1} register on the interfer radio, as shown in red in the bottom of Fig. \ref{fig:interference}.  As the interferer's power increases, the samples of quantized RSS begin to take two different RSS values, both -54 and -53 dBm.  The probability of -54 (and -53) changes as a function of whether the person is inhaling or exhaling. This then allows estimating the periodicity of the signal. This is shown to enhance the accuracy of breathing rate estimation by lowering the RMSE of breathing rate estimation from 15 bpm to 2 bpm. For the given data, we also note that increasing the power of the interference beyond a certain point, will not improve accuracy below 2 bpm of RMSE.

%% file: crb.tex
\section{Bounds for RSS Surveillance}
\label{sec:bound}

The previous section provides experimental evidence of the possible benefits of HI, which an attacker can exploit to improve performance when RSS is quantized.
While the experimental results provide examples of what an attacker could achieve, they do not provide any guarantees about the best performance an attacker could achieve. In this section, we provide analytical limits on the attacker's eavesdropping capability.  These limits consider that an attacker may use HI, and are a function of the system parameters of available RSS sampling rate and quantization step size, as well as the amplitude of the breathing signal being surveilled by the attacker. As before, we use the variance of breathing rate estimates as an example.  Note however that the bound is generally applicable to any RF sensing application which estimates the amplitude or frequency of a sinusoidal signal component. We compute the theoretical lower bounds on estimation variance using Cram\'er-Rao bound analysis. 

The bounds on variance provide guarantees that are useful to both users and RFIC system designers.  First, note that one can never guarantee that an attacker cannot estimate breathing rate at all --- an attacker can always estimate breathing rate to be 15 bpm, for example, without any RSS data, but this would not be a useful attack.  We focus on bounding the lowest possible variance of an attacker's unbiased breathing rate estimate since, if this variance is high, it effectively shows that the attacker is unable to gain meaningful information about the person's breathing rate.  A user could use such a bound to decide if an attacker who compromises the device could effectively estimate their breathing rate or not.  An RFIC designer could alter the parameters of the RSS measurements made available from the chip to increase the variance bound and thus make their device more acceptable to privacy-conscious customers.

\subsection{RSS Model for Breathing}
\label{ssec:model} 

In order to derive the theoretical bounds on breathing eavesdropping, we first model the received power including the variation due to breathing. Published work has demonstrated the validity of modelling of the breathing-induced component in the received power as a sine wave \cite{patwari2014monitoring}. Thus, we assume that a person's breathing changes the measured signal as an additive sinusoidal component.  Here we again use \emph{received power} to indicate the continuous-valued power of the signal at the antenna, and RSS to indicate the quantized discrete-valued power reported by the transceiver IC.  Although generally an eavesdropper may take burst measurements, we assume a scalar time-dependant signal for simplicity.  We use $B$ to denote the received power when breathing is absent, and $v(k)$ to denote the noise in sample $k$, which is assumed to be zero-mean white Gaussian noise with variance $\sigma^2$ \cite{yiugitler2017log}.  Therefore, the sampled received power signal is given as
\begin{equation} \label{eqn:model}
    x[k] = A\cos{(\omega T_s k+\phi)}+B + v[k],  \qquad k \in \mathbb{Z},
\end{equation}
where $T_s$ is the sampling period, and the breathing signal has unknown amplitude $A$, DC offset $B$, frequency $\omega$, and the initial phase $\phi$. The unknown parameter vector is  $\bm{\theta} = [A, B,\omega,\phi]^T$.  

Our model is that the received power is quantized with a step size of $\Delta$.  Typically $\Delta \gg A$, that is, the step size is significantly larger than the changes in RSS due to breathing. Eexperimental work has indicated peak-to-peak amplitudes of the breathing signal to be, for example, 0.1 dB \cite{luong2016rss}, or 0.3 dB \cite{uysal2017contactless}. Pulse-induced amplitudes are even smaller.  It is rare to see transceiver RSS to be quantized to less than 0.5 dB, as typical step sizes are 1.0 dB or higher.  Since $A$ is low compared to $\Delta$, during breathing monitoring, the (quantized) RSS measurement typically takes one of two neighboring values.  It follows that we can approximate the RSS signal as a one-bit quantization of the received power $x[k]$. 
Note this approximation is not imperative for obtaining an estimation bound, since multi-bit CRB analysis of frequency analysis is possible \cite{moschitta2007cramer}.  When $\Delta \gg A$, that more complicated bound is nearly identical to the bound we derive, but the complexity can obscure the lessons learned from the bound with the one-bit quantization assumption.

Assuming one-bit quantization, the RSS is represented as:
\begin{equation} \label{eqn:quantization}
y[k] = {\rm sign}\left(x[k]-\zeta\right),
\end{equation}
where $\zeta$ is the threshold for quantization (the boundary between the two RSS bins) and the sign function ${\rm sign}(\cdot)$ is defined as ${\rm sign}(x)=1$ for $x\geq0$ and ${\rm sign}(x)=-1$ for $x < 0$. Without loss of generality, we assume $\zeta=0$.  The DC offset $B$ becomes the distance from the nearest quantization threshold and takes a value in the set $[-\Delta/2, \Delta/2]$. We define ${\bm y} = \left[y[0], \ldots, y[N-1]\right]^T$ to be our measurement vector.  

The attacker's goal is to estimate $A$ and $\omega$ from these RSS measurements ${\bm y}$. 
While rate is most frequently discussed in breathing monitoring,  
the amplitude $A$ is also useful as it may indicate breathing volume \cite{nguyen2016continuous}, which then may indicate the person's level of activities or stress level. It is also a useful parameter in detecting the presence of a breathing person near a link. 

\subsection{Cram\'er-Rao Bound (CRB) Analysis}
\label{ssec:CRB}
Here, we compute the Cram\'er-Rao bound of the parameter vector ${\bm \theta}$ given the measurements ${\bm y}$. First we define the probability mass function corresponding to $k^{\mbox{th}}$ sample $y[k]$ as 
\[
f_{y[k]}=P(y[k]=q ; \, {\bm \theta}), \qquad q \in \{-1, +1\}.
\]
If we define ${\cal C}_k \coloneqq \cos (\omega T_s k+\phi)$ and ${\cal S}_k \coloneqq \sin (\omega T_s k+\phi)$, then
\begin{equation}
\begin{aligned}
f_{y[k]}(q;{\bm \theta}) &= P(y[k]=q;{\bm \theta}) =P(qx[k]>0;{\bm \theta})  \\ 
&= \frac{1}{\sqrt{2\pi}\sigma}\int^{\infty}_0 \exp\left({-\frac{[x-q(A{\cal C}_k+B)]^2}{2\sigma^2}}\right)dx.
\end{aligned} \nonumber
\end{equation}
Equivalently,
\begin{equation}
\begin{aligned}
f_{y[k]}(q;{\bm \theta})= \dfrac{1}{2}{\rm erfc}\left(-\dfrac{q}{\sqrt{2}\sigma}(A{\cal C}_k +B)\right).
\end{aligned}
\end{equation}
To compute the CRB, we first derive the Fisher information matrix (FIM).
From \cite{kay}, the element of the FIM from $i^{th}$ row and $j^{th}$ column is given by
\begin{equation}
\begin{aligned}
{\bm I}(\bm{\theta})_{ij} &= {\rm E}\left[  \dfrac{\partial \log f_{{\bm y}}(\bm{q};{\bm \theta})}{\partial \theta_i} \dfrac{\partial \log f_{{\bm y}}(\bm{q};{\bm \theta})}{\partial \theta_j}\right]\\
\end{aligned}
\end{equation}
Since the variables $y[k]$ are independent, the element of the FIM from $i^{th}$ row and $j^{th}$ column becomes:
\begin{equation} \label{eqn:fim_our_case}
\begin{aligned}
{\bm I}(\bm{\theta})_{ij} &= \sum_{k=0}^{N-1}\sum_{q=\pm 1} \frac{1}{f_{y[k]}(q;{\bm \theta})} \dfrac{\partial f_{y[k]}(q;{\bm \theta})}{\partial { \theta_i}} {\dfrac{\partial f_{y[k]}(q;{\bm \theta})}{\partial {\theta_j}}}
\end{aligned}
\end{equation}
We compute the partial derivatives for the parameters $\bm \theta$, plug them into (\ref{eqn:fim_our_case}), and the resulting FIM becomes:

\begin{equation}
\label{eqn:fim}
{\bm I}(\bm{\theta}) = \frac{2}{\pi \sigma^2} \sum_{k=0}^{N-1}
\dfrac{\exp{\left(-\frac{1}{\sigma^2}\left(A{\cal C}_k+B\right)^2\right)}}{1- {\rm erf}^2\left(\frac{1}{\sqrt{2}\sigma}(A{\cal C}_k +B)\right)}
{\bm F}_k,
\end{equation}
where

\begin{equation}
{\bm F }_k= 
\begin{bmatrix} 
{\cal C}_k^2  & {\cal C}_k & -AkT_s {\cal S}_k{\cal C}_k  & -A{\cal S }_k{\cal C}_k 
\\
{\cal C}_k    & 1        & -AkT_s {\cal S}_k   & -A{\cal S}_k 
\\
-AkT_s {\cal S}_k{\cal C}_k  & -AkT_s {\cal S}_k   & A^2 k^2T_s^2 {\cal S}_k^2  & A^2kT_s {\cal S}_k^2
\\
-A{\cal S}_k{\cal C}_k & -A{\cal S}_k & A^2  kT_s{\cal S}_k^2  &  A^2{\cal S}_k^2 
\end{bmatrix}. \nonumber
\end{equation}

In this analysis, we focus on finding the bounds on variance of unbiased amplitude ($\hat{A}$) and frequency ($\hat{\omega}$) estimators, which are given in the inverse of the FIM in (\ref{eqn:fim}) as
\begin{equation}
\begin{aligned}
\text {CRB}(\hat{A})&=\left\{{\bm I}(\bm{\theta})^{-1}\right\}_{11},  \\ 
\text {CRB}(\hat{\omega})&=\left\{{\bm I}(\bm{\theta})^{-1}\right\}_{33}.
\end{aligned}
\end{equation}

For a quantization step size $\Delta$, the DC offset $B$ can be represented as a uniform random variable defined over the interval $[-\Delta/2,\Delta/2]$. In addition, each bound is a weak function of the initial phase $\phi$ which is also random and uniform for our applications. Thus we average the CRB over uniform phase and uniform DC offset.  
We use $\overline{ \text {CRB}}$ to indicate the CRB averaged over a uniform phase $\phi$ and uniform DC offset $B$. Therefore, the variance of amplitude estimates $var(\hat{A})$ and the variance of frequency estimates $var(\hat{\omega})$ are bounded by $\overline{\text {CRB}}(\hat{A})$ and $\overline{\text {CRB}}(\hat{\omega})$ respectively.
\begin{equation}
\label{eqn:bound}
\begin{aligned}
var(\hat{A})&\geq \overline{\text {CRB}}(\hat{A})\\ 
var(\hat{\omega})&\geq \overline {\text {CRB}}(\hat{\omega}).
\end{aligned}
\end{equation}

In subsequent subsections, we study the accuracy of breathing eavesdropping based on estimation variance computed in (\ref{eqn:bound}). In particular, we analyze the effects of quantization step size $\Delta$, interference $\sigma$ and sampling rate $f_s$ .

\subsection{Effects of Helpful Interference}
\label{ssec:noise}
We study the effect of adding noise to RSS measurements prior to quantization for both amplitude and frequency estimation. For this analysis, we consider a value of frequency of  $f$ = 15 bpm, i.e., 0.25~Hz, which is a typical resting breathing rate for a normal adult.  Further, we consider a sampling rate $f_s$ = 10~Hz. We set $N$, the number of samples, such that $NT_s = 30$ s, and we choose an amplitude $A=0.1$ dB.

\begin{figure*}[tbph]
  \begin{center}
  
     \includegraphics[width=0.4\textwidth]{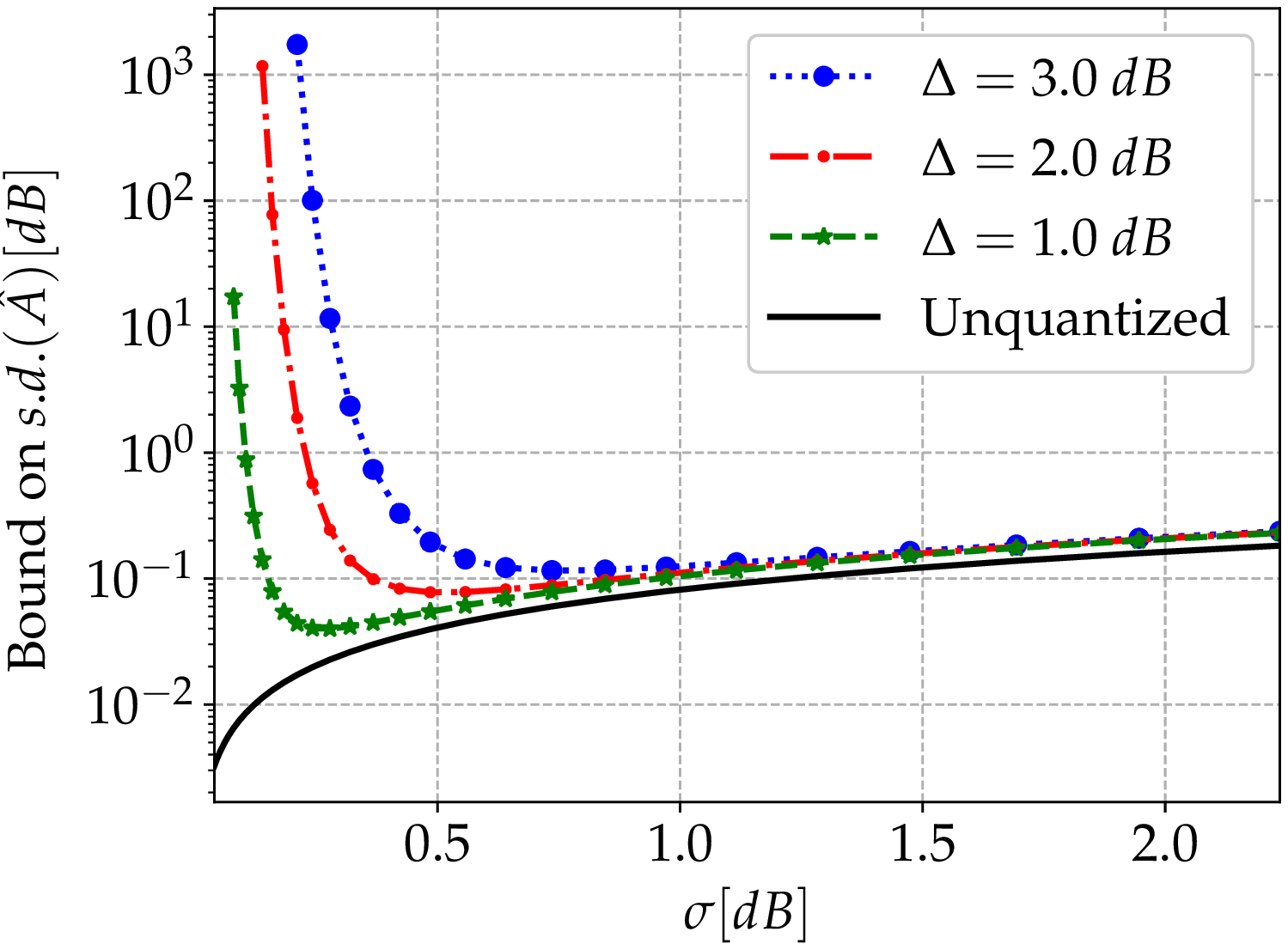}
$\,\,\,\,\,\,\,\,$
     \includegraphics[width=0.4\textwidth]{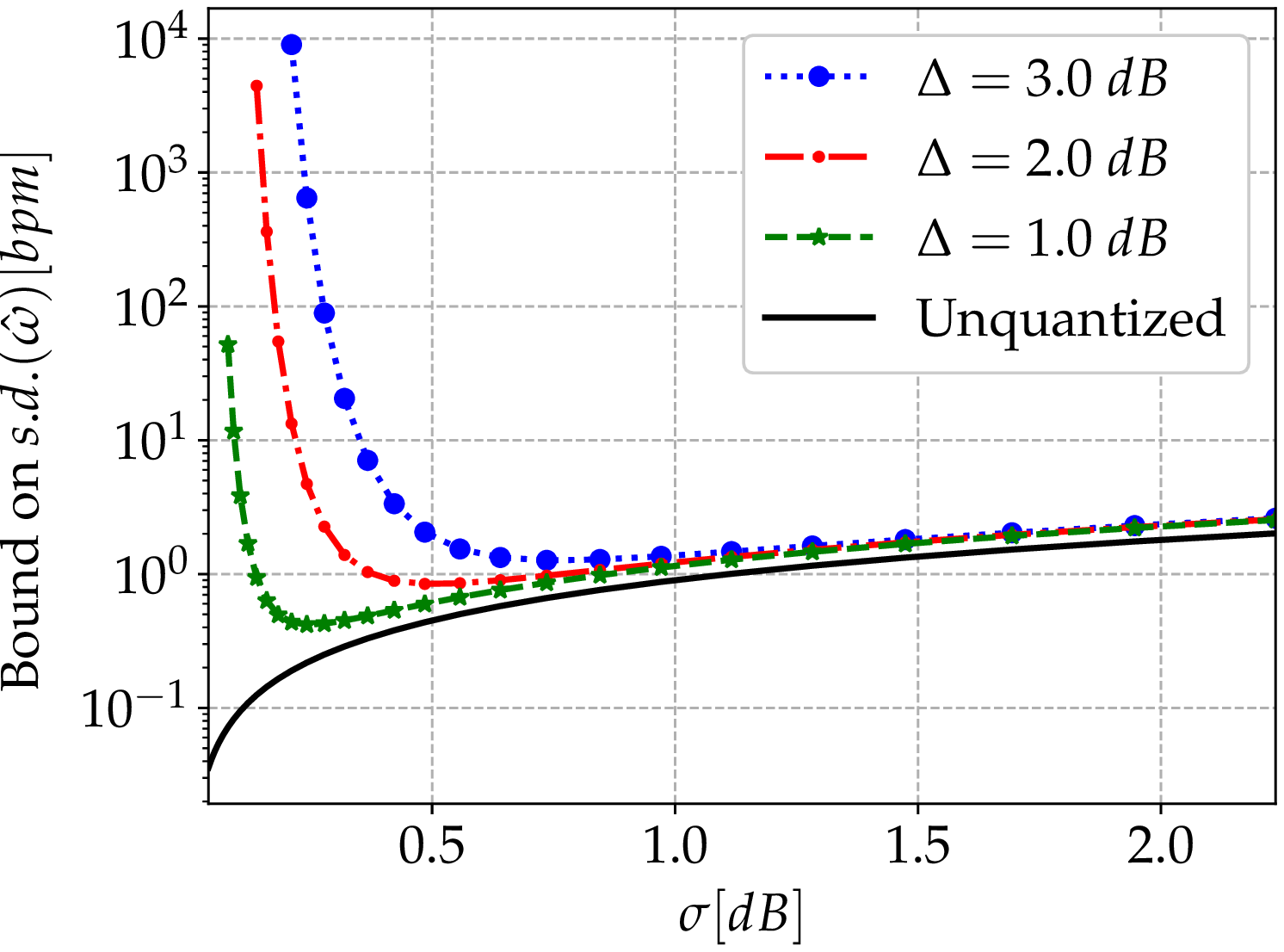}
\end{center}
  \caption{CRB of (Left column) amplitude $A$ and (Right column) frequency $\omega$, vs.\ noise power when $f_s = 10~Hz$. As noise power increases, the estimation variance decreases and then slowly increases.} \label{fig:noise_crb}
 \end{figure*}

We plot numerical results in Fig. \ref{fig:noise_crb} for the bound on the standard deviation of amplitude estimates as a function noise standard deviation as computed from (\ref{eqn:bound}).
We note that as the noise power increases, the bound on standard deviation of amplitude estimate generally decreases for quantized RSS measurements. Intuitively, this is because, as the sine wave is more likely further away from the threshold, even at its maxima or minima, estimation accuracy requires higher noise power in order to ensure that the measurements are not purely from one quantization level. For quantized RSS,  small interference power results in higher estimation variance as the quantized RSS measurements have a lower probability of changing their RSS levels with small noise power. 

This effect is similarly observed in frequency estimation. Fig. \ref{fig:noise_crb}(right column) shows the effects of increasing noise power to RSS prior to quantization on the accuracy of breathing rate estimation. We see that increasing the noise level decreases the estimation variance for quantized RSS measurements. These results indicate that adding HI to a wireless channel improves the accuracy of amplitude and frequency estimations from quantized RSS measurements.  They also match the characteristics observed experimentally in \S \ref{sec:eval}.

It is also worthwhile to see the effect of adding noise when the signal is not quantized. We see the bound for $\hat{A}$ from \cite{rife1974single}, is $\mbox{var}(\hat{A}) \ge 2\sigma^2/N$, which indicates that the standard deviation increases with noise power for unquantized RSS measurements. Similarily,  we see that increasing the noise level increases the estimation variance of frequency estimates for unquantized RSS measurements. 

\vspace{0.1in}

\begin{figure}[h]
  \begin{center}
  \includegraphics[width=0.9\columnwidth]{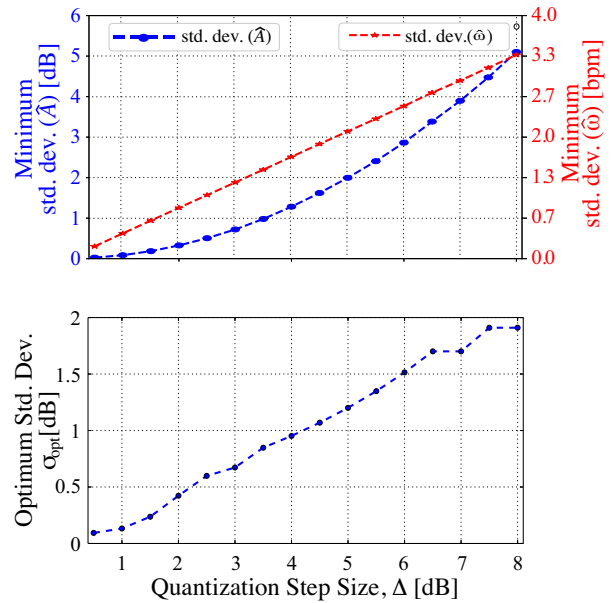}
   \end{center}
  \caption{CRB as a function of quantization step size}\label{fig:stepsize_bound}
 \end{figure}

\noindent  {\bf Optimum Noise Variance}:
A key observation from the results is that the bound on estimation variance using quantized measurements has a minimum value with respect to noise power for a given sampling rate and quantization step size. 
We note from Fig. \ref{fig:noise_crb} that there exists an optimal noise level at which estimation variance is brought to its minimum for a given sampling rate and RSS quantization step size. 
Our numerical results show that the optimal noise level for amplitude estimation matches that for frequency estimation. In Fig. \ref{fig:stepsize_bound}, we observe that the standard deviation of this optimal noise is linearly proportional to the RSS quantization step size, and that $\sigma_{opt}$ is approximately $\Delta / 4$. 
Interestingly, this standard deviation of helpful interference is just less than the standard deviation of quantization error, which is $\Delta/\sqrt{12} = \Delta / 3.46$.

\begin{figure}[btph]
\begin{minipage}[h]{0.95\linewidth}
\centerline{\includegraphics[width=0.95\columnwidth]{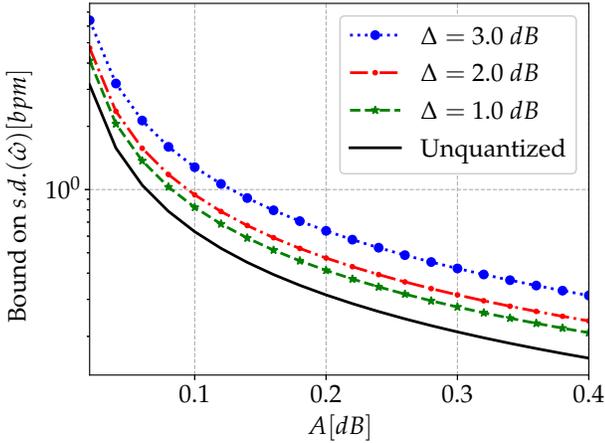}}
\end{minipage}
\caption{Effect of amplitude on frequency estimation}
\label{fig:ampf}
\end{figure}

It should be noted that the bound on standard deviation of $\hat{\omega}$ is a weak function of the frequency parameter $\omega$, and thus the plot is omitted. However, the amplitude significantly affect the performance of frequency estimation; as shown in Fig. \ref{fig:ampf}, higher amplitude results in lower standard deviation of frequency estimates.
 
 \subsection{Effects of Quantization Step Size}
An other parameter that controls the performance of breathing eavesdropping is the quantization step size $\Delta$. 

We can see that the accuracy of breathing eavesdropping, despite the ability of the attacker to use helpful interference, can be generally be degraded by increasing the RSS quantization step size. Furthermore, the minimum estimation bounds for amplitude and frequency estimates behave differently with respect to the RSS quantization step size. In Fig. \ref{fig:stepsize_bound}(bottom), we observe that the bound for frequency estimates increases linearly with RSS quantization step size whereas the bound for amplitude estimate fits a quadratic model with respect to the step size.

\begin{figure*}[phtb]
\begin{center}
     \includegraphics[width=0.4\textwidth]{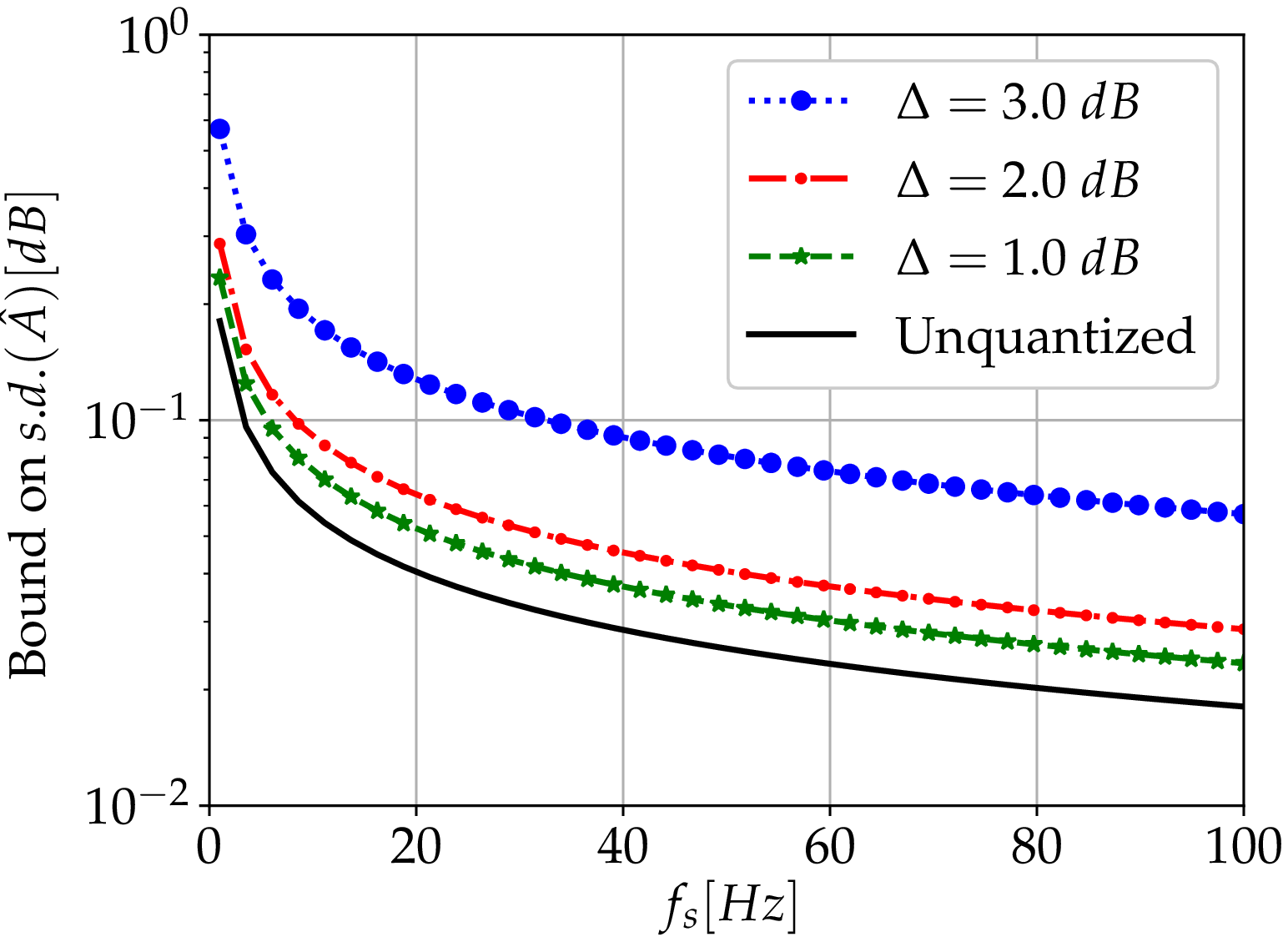}
$\,\,\,\,\,\,\,\,$
     \includegraphics[width=0.4\textwidth]{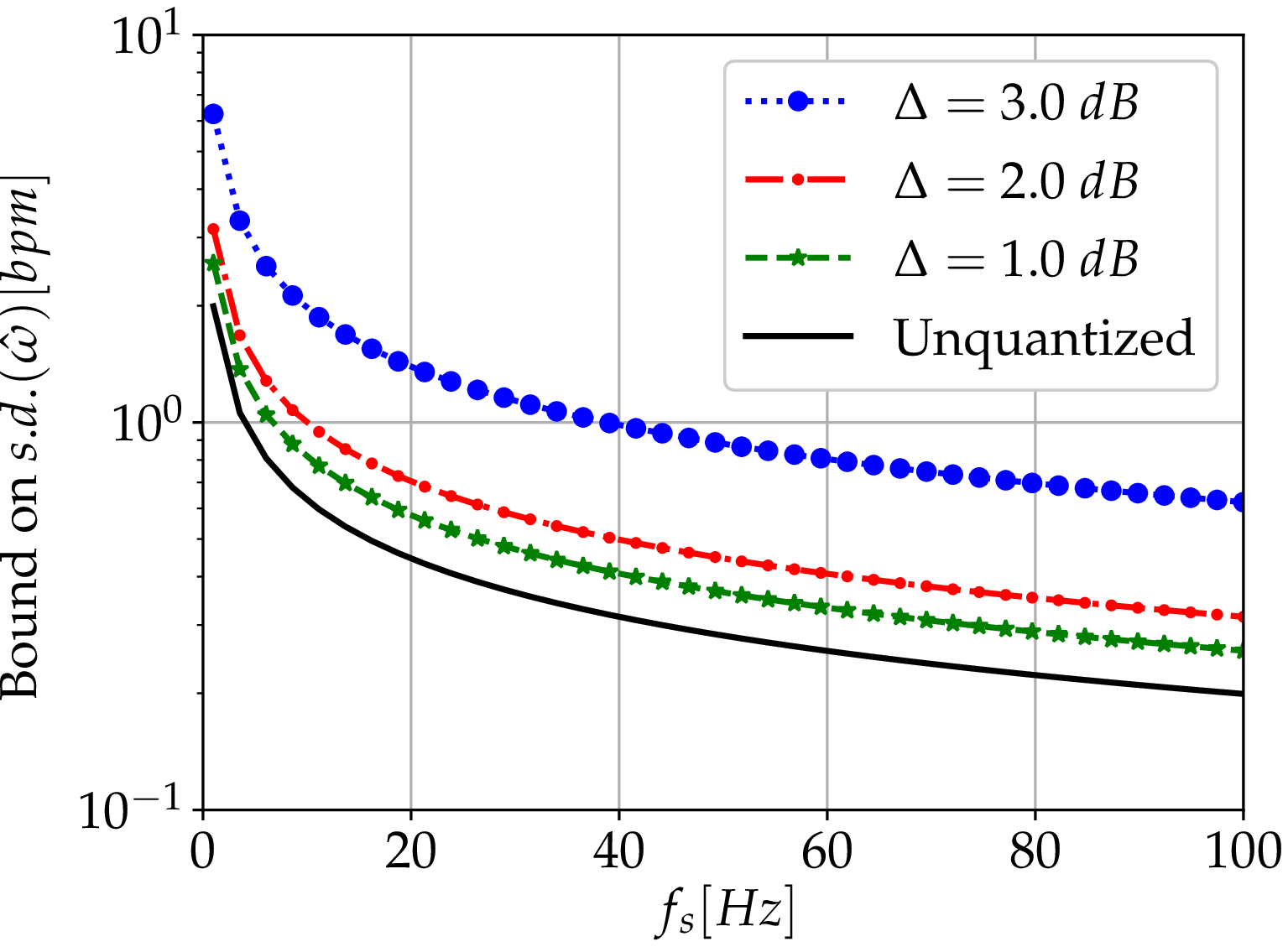}
\end{center}
  \caption{CRB of (Left column) amplitude $A$ and (Right column) frequency $\omega$, as a function of sampling rate when $\sigma = 0.7~dB$. As sampling rate $f_s$ increases, the estimation variance decreases.}
 \label{fig:sample_crb}
 \end{figure*}

\subsection{Effects of RSS Oversampling}
\label{ssec:oversample}
Next, we evaluate the effects of sampling rate on the accuracy breathing monitoring, particularly in breathing rate estimation. We use $\omega$ = 1.55~rad/s which corresponds to $f = 0.25$ Hz $= 15$ bpm, and $A=0.1$ dB. 
In Fig.~\ref{fig:sample_crb}(Left column), we plot the bound on standard deviation of amplitude estimate as a function of the sampling rate $f_s$. This bound decreases monotonically with $f_s$ for any value of the quantization step size $\Delta$.  The lowest $f_s$ in Fig.~\ref{fig:sample_crb}(Left column) is 1 Hz.  This suggests that an eavesdropper attains lower estimation variance by collecting RSS at higher rate. If it is possible to increase the sampling rate, the bound shows the possibility of order-of-magnitude decreases in standard deviation. Similar results are observed for frequency estimation where increasing sampling rate decreases the bound on standard deviation of frequency estimates.

\subsection{Overall Effects}
\begin{figure*}[tbph]
  \begin{center}
  
     \includegraphics[width=0.4\textwidth]{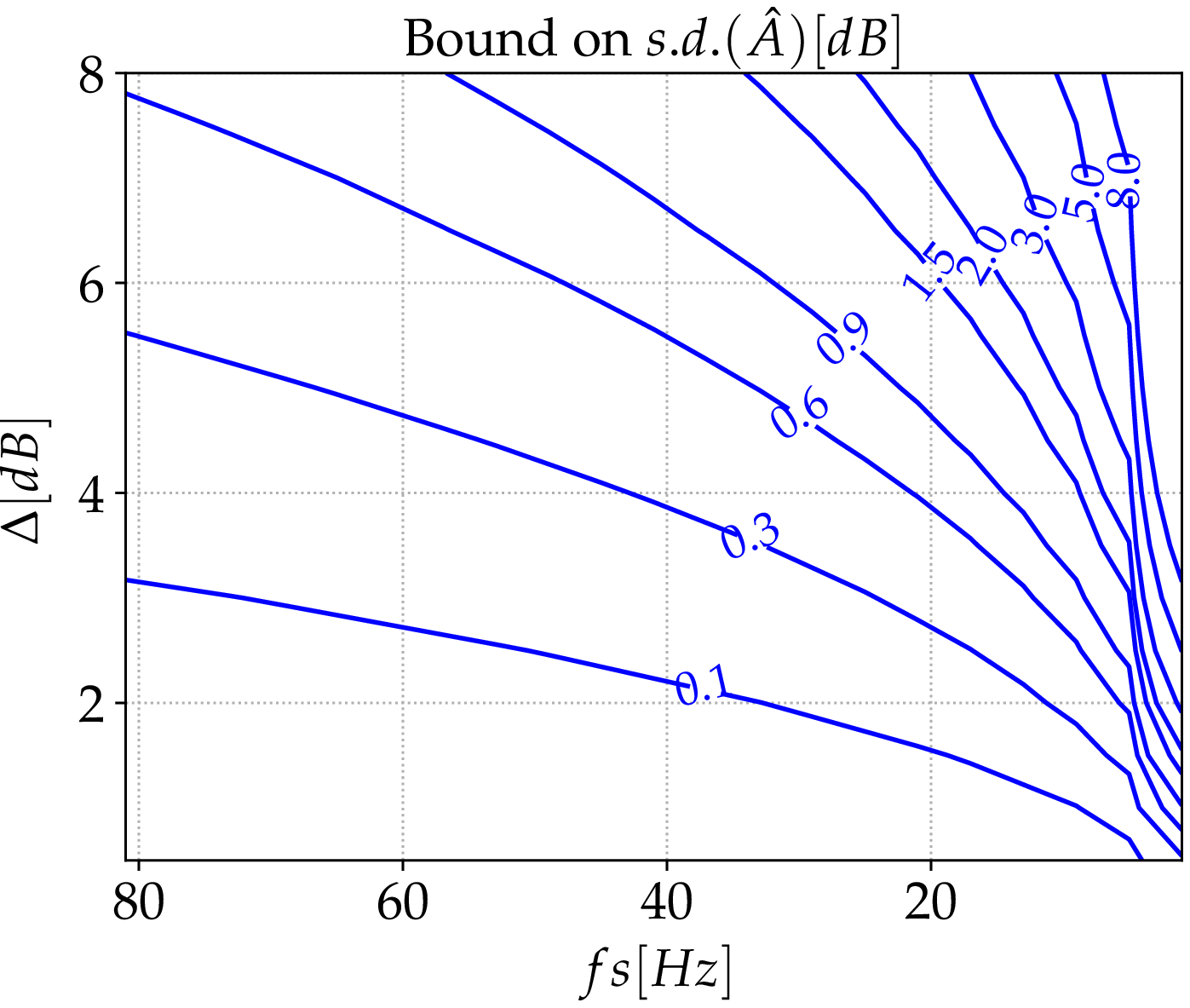}
$\,\,\,\,\,\,\,\,$
     \includegraphics[width=0.4\textwidth]{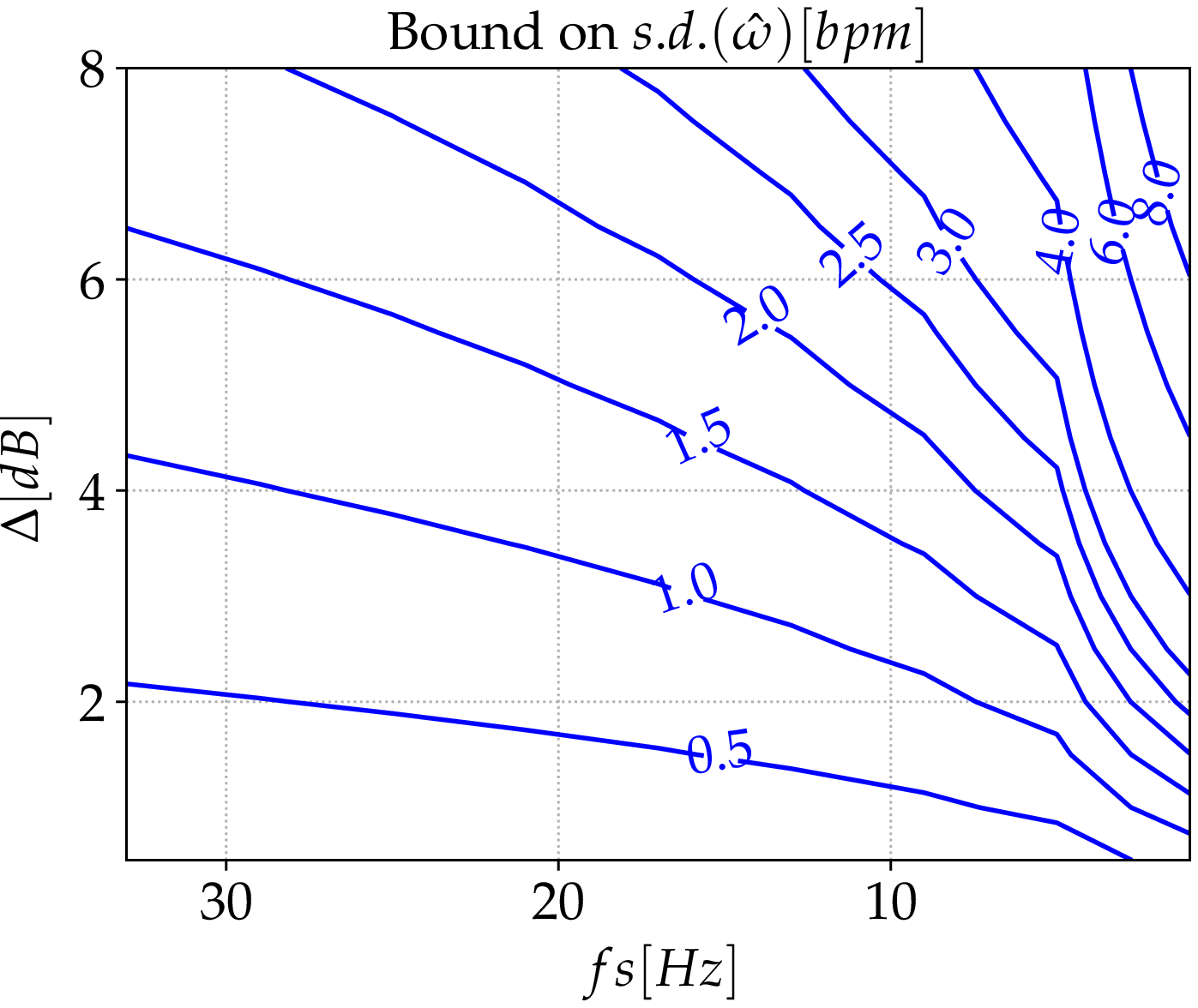}

\end{center}

  \caption{Contour plot of minimum CRB for (Left column) amplitude and (Right column) frequency estimates vs.\ sampling rate and quantization level.} \label{fig:sample_noise}
 \end{figure*}

Our CRB analysis shows that the accuracy of RSS-based surveillance can be controlled mainly by two parameters: RSS quantization step size and sampling rate at which the RSS measurement is collected.  
We numerically analyze the combined effect of oversampling and quantization on RSS-based breathing monitoring. Fig. \ref{fig:sample_noise}(Left column) shows the minimum theoretically achievable standard deviations of amplitude and frequency estimates as functions of quantization step size $\Delta$ and RSS sampling rate $f_s$. We note from these plots that lower estimation variance is generally attained with higher sampling rate and low step size. On the other hand, lower sampling rate and high step size lead to large estimation variance and hence lower accuracy in breathing surveillance. These results suggest the accuracy of an RSS-based breathing surveillance attack can be controlled by a designer by selecting a large RSS step size and low sampling rate.

%% file: mitigation.tex
\section{Mitigation}
\label{sec:mitigation}

The numerical results from the bound analysis presented in Section \ref{sec:bound} are, frankly, quite discouraging. 
Existing wireless systems generally provide access to the RSS at a rate of up to 1000s of times per second, and at quantization step sizes as low as 0.5 dB. Even with two orders of magnitude lower sampling rate, and a quantization step size of 2 dB, such devices, if compromised, would enable an attacker to know breathing rate within a standard deviation of just under 1 breath per minute.  In other words, they are inherently vulnerable to RSS-based breathing surveillance. 

Clearly, there is a need to design systems that limit RSS information in new ways.  In this section, we present a group of mitigation strategies that could be used by system developers and RFIC designers to make their devices more secure against an RSS-based surveillance attack, without significantly compromising the performance of wireless systems.

{\bf Less RSS Information:} Ideally, RSS-based eavesdropping can be deterred by limiting access to RSS information from the RFIC with low sampling rate and very large quantization step size. However, this approach may be impractical because the effectiveness of almost any mobile wireless protocol depends on accurate RSS information. If we are unable to access RSS, we cannot use power control to reduce energy consumption and limit interference, for example.

{\bf Never Provide Both:} A more realistic recommendation would be that the transceiver IC could be limited to prevent \emph{both} high rate and low step size RSS measurements as follows. In this scheme, applications which require accurate RSS are provided with high resolution RSS at a low rate, while applications demanding RSS at a high rate would be allowed RSS only with large quantization step size. Only one application could be active at a time, and there would be a long switching time whenever the setting was switched.  In Fig. \ref{fig:sample_noise}, we see theoretically minimum deviation for breathing rate estimated from RSS measurements, as a function of RSS step size and sampling rate. Normally, clinical-grade breathing monitors have an accuracy of breathing rate within 1 bpm \cite{clifton2007measurement}. Therefore, a privacy-aware system developer may want the minimum standard deviation of breathing rate estimated by eavesdropper to be 2 bpm. In this case, the developer achieves this requirements by designing a wireless system with a parameter pair of a sampling rate and RSS step size picked anywhere above the 2 bpm curve in Fig. \ref{fig:sample_noise}, for example, at 20 Hz with a step size of 8 dB.  The low rate setting could allow RSS at 4 Hz with a step size of 2 dB.  

However, this may not be appropriate for mobile communications devices with tight transmit power control requirements.  For example, 4G and other code-division multiple (CDMA) systems are limited by the near-far problem, which for mobile devices requires a few hundred Hz of power control updates.  This high rate is due to the high rate at which the mobile channel may be changing.  For this application 1 dB, equivalently 25\% of the power in linear terms, is a significant step size for CDMA receivers; and as a result low step size is required.

{\bf Adaptive Rate Scheme:}  For such applications, we suggest an adaptive rate RSS reporting mechanism in which applications are granted high quality RSS at very high rate only when the wireless channel is changing rapidly. When the channel is nearly constant, new RSS measurements are provided only at a low rate, e.g., less than 0.25 Hz.  Note that when the channel is changing rapidly, it is either due to the mobility of the endpoints of the link, or fast motion of people near the link.  In either case, breathing is largely unable to be estimated with mobile endpoints or with fast human motion in the vicinity of the link \cite{adib2015smart}.  Thus we do not anticipate breathing monitoring being possible from devices with this adaptive rate RSS scheme.

{\bf Adaptive Quantization Scheme:}  We observe that when the offset $B$ (distance from the average RSS to the nearest quantization threshold) is large, breathing monitoring is most difficult.  We can ensure that $B$ is large by changing the quantization boundaries.  In this scheme, we would have two quantization functions with the same step size $\Delta$.  For example the first quantization scheme would have boundaries $\{RSS_0 + n \Delta\}$ for some lowest threshold $RSS_0$ and non-negative integers $n=0, 1, \ldots$, while the 2nd quantization scheme would have boundaries $\{ RSS_0 + (n+0.5) \Delta \}$.  Each period of time, e.g., 10 or 20 seconds, the quantization scheme would be re-selected to maximize $B$.  Since $A \ll B$, during periods of time when RSS could be used for breathing monitoring, since $B$ was large, there would primarily be only one value of RSS reported for long periods of time.  This scheme might also be used in combination with either of the previous two schemes.

%% file: related.tex
\section{Related work}
\label{sec:related}

{\vspace{0.1in} \noindent \bf Radio Frequency Sensing:}  
Radio Frequency sensing is the use of radio channel measurements to monitor vital signs or detect human activity without putting any sensor on the body of a person. 
Various radio channel measurements such as received signal strength (RSS), channel impulse response (CIR), and channel state information (CSI) have been used for multiple RF sensing applications including contact-free vital sign monitoring \cite{patwari2014breathfinding,liu2015tracking}, device-free localization \cite{kaltiokallio2012follow, yang2013rssi}, gesture and activity recognition \cite{wang2015understanding, sigg2014telepathic}, and human identification \cite{xin2016freesense}. 

Among several channel measurements employed in most commercial wireless systems, RSS is considered to be the most widely available measurement across diverse wireless platforms \cite{liu2017wireless}.
RSS has been applied in various RF sensing applications including device-free localization \cite{youssef2007challenges}, contact-free breathing monitoring \cite{kaltiokallio2014non, patwari2014breathfinding}, security \& surveillance \cite{hussain2008using}, activity and gesture recognition \cite{sigg2014rf,luong2016rss}, and home monitoring \cite{kaltiokallio2012follow}. 

The ease of access to RSS and its capability in RF sensing allows a potential threat on privacy. Little attention has been paid to these threats, mainly because RSS-based sensing has been reported to have limited reliability as a result of its relatively large quantization step size \cite{luong2016rss}. However, the limits on the capability of RSS-based sensing has not been fully explored. In this paper, we experimentally demonstrate such an unexplored RF sensing capability that uses noise superimposed in RSS measurements to improve breathing monitoring.

{\vspace{0.1in} \noindent \bf Estimation Bounds:}  
Estimation bounds are statistical tools used to evaluate the performance of algorithms in estimating certain parameters of interest with respect to the maximum theoretically attainable accuracy, commonly based on their estimation variance. 

The Cram\'er Rao lower bound is the most common variance bound due to its simplicity \cite{kay}.  It provides the lowest possible estimation variance achieved by any unbiased estimator.

Evaluating the accuracy of algorithms used in RSS surveillance is essential step to determine eavesdropper's capabilities.
An analytical explanation for the relationship between noise power, sampling rate, amplitude, quantization, and parameter estimator performance has not been presented. 
The change in RSS due to periodic activities like respiration can be generally modelled as a sine wave \cite{patwari2014monitoring}.  
For unquantized sine wave signal, the CRB on the variance of unbiased frequency estimators is derived in \cite{kay, rife1974single}. 
However, RSS has a signficantly higher quantization step size than the typical RSS change induced by vital signs, each RSS sample primarily falls into either of two successive RSS values. 
H{\o}st-Madsen \emph{et al.} \cite{host2000effects} quantitatively explain the effect of quantization and sampling on frequency estimation of a one-bit quantized complex sinusoid, 
but without presenting bounds for amplitude estimation or considering a DC offset as a complicating parameter.  In this paper, we evaluate attacker's bound on breathing rate and amplitude estimation in the realistic case that there is a DC offset.  Further we have demonstrated what is observed in the bound, that increased interference power can actually help improve estimates.

{\vspace{0.1in} \noindent \bf Wireless Network Security:}  
Security in wireless networks is conventionally achieved through cryptographic protocols at multiple layers in the network stack. In wireless local area networks including 802.11, a number of cryptographic protocols have been standardized including IPSecurity, 
 Wi-Fi Protected Access (WPA), and Secure Sockets Layer (SSL). 
Due to the broadcast nature of the wireless medium, researchers have proposed additional security protocols at the physical layer to deter eavesdropping and jamming, such as by exploiting channel characteristics \cite{tomko2006physical, li2005mimo}, employing coding schemes \cite{shiu2011physical}, or controlling signal power \cite{noubir2004connectivity,gollakota2011they}. 
However, these approaches do not prevent an adversary already with some access to a system from using PHY layer signal measurements for sensing purposes.

Such an adversary can also force a wireless device to continuously transmit helpful interference in order to reduce the effect of quantization on the quality of the RSS information. Most wireless standards use a multiple access control method to avoid interference, such as carrier-sense multiple access (CSMA). However, many RFICs (e.g., Atheros AR9271) provide the option to disable CSMA and control the random backoff timer \cite{vanhoef2014advanced}. These vulnerabilities pave the way for the attacker to change a device's software to create an interferer operating on the same channel at the same time as the receiver measuring RSS.

Despite considerable research effort for privacy, RSS surveillance attack by exploiting measurements from wireless systems has been unresolved problem. 
Banerjee \emph{et al.} demonstrate  that an attacker can easily estimate artificial changes in transmit power to detect and locate people through wall. In \cite{qiao2016phycloak}, an third device is introduced to distort the PHY layer signal before it could be used by eavesdropper for sensing purposes, but fails for multiple-antenna eavesdropper or if a device can be remotely compromised and caused to run the attacker's software.

%% file: discussion.tex
\section{Discussion}
\label{sec:future}

Recent progress in RF sensing has demonstrated the ability to re-purpose standard wireless devices for vital sign monitoring.  From the perspective of a privacy-conscious user, this can be considered a significant threat.  A person's smart home, if hacked, could provide an attacker constant track of their vital signs, their mood and activities, and perhaps record them and be able to guess what they're typing.  While it may be obvious to a user that smart home \emph{sensors} such as video cameras are privacy risks, it may not be so obvious that standard transceivers may also pose such risks.   

From another perspective, the attack we present is more readily addressed.  By reducing access to RSS measurements from the transceiver, perhaps by implementing one of the schemes we describe in \S \ref{sec:mitigation}, RFIC or SoC designers can make transceivers which are \emph{unable} to be useful in the described breathing monitoring attack. The attack presented in \cite{qiao2016phycloak} cannot readily be prevented since attackers may bring to the area their own devices with potentially unlimited measurement capabilities.  However, device manufacturers would need to be convinced that consumers care about privacy from RF sensing in order to want to change how their devices provide access to RSS.

This paper uses breathing monitoring as the main example, however, the same mathematical framework applies to the surveillance of other periodic signals such as pulse rate. The pulse-induced signal in received power usually has a lower amplitude and higher frequency than the breathing-induced signal \cite{adib2015smart}. As a result, mitigation strategies that prevent remote breathing monitoring also protect privacy against pulse rate monitoring.

Future work should study limits on the capability of a surveillance attack on gestures and activities from RSS measurements. In gesture, activity, and keystroke recognition, spectral and time-frequency measures are often used as features input to machine learning algorithms \cite{ali2015keystroke}. Current bound results on amplitude and frequency estimation could be extended to study an attacker's capability in RSS based gesture and activity recognition.

%% file: conclusion.tex
\section{Conclusion}
\label{sec:conc}

In this paper, we explore the limits on RSS-based surveillance attacks.  We analyze the capability of such attacks in estimating the sinusoidal parameters of low-amplitude sinusoidal signals by deriving the theoretical lower bound with which an attacker could estimate the rate and amplitude of a sinusoid.  We show, both theoretically and experimentally, that the adversary could force other wireless devices to transmit simultaneously in order to improve their estimates.  
The numerical values of the lower bound on variance show, for typical RFICs, an RSS-surveillance attack could be very accurate.
We discuss, as a result, how a device designer could limit the performance of a potential attack by adjusting the quantization step size and the sampling rate. Most commercial transceivers have fixed RSS quantization schemes, however, a manufacturer could use adaptive quantization methods to ensure that breathing monitoring attacks are ineffective.

%% file: ack.tex
\section*{Acknowledgement}
This material is based upon work supported by the US Army Research Office under Grant No. 69215CS.